\documentclass[prl,twocolumn,aps]{revtex4}
\usepackage{verbatim}
\usepackage{graphicx}
\normalfont
\begin{document}


\title{Voltage controlled nuclear polarization switching in a single InGaAs quantum dot}
\author{M. N. Makhonin$^{1}$, J. Skiba-Szymanska$^{1}$, M. S. Skolnick$^{1}$, H.-Y. Liu$^{2}$, M. Hopkinson$^2$, A. I. Tartakovskii$^{1}$}
\address{$^{1}$ Department of Physics and Astronomy, University of Sheffield, S3 7RH,UK\\
$^{2}$ Department of Electronic and Electrical Engineering, University of Sheffield, Sheffield S1 3JD, UK}

\date{\today}

\begin{abstract}
Sharp threshold-like transitions between two stable nuclear spin polarizations are observed in optically pumped individual InGaAs self-assembled quantum dots embedded in a Schottky diode when the bias applied to the diode is tuned. The abrupt transitions lead to the switching of the Overhauser field in the dot by up to 3 Tesla. The bias-dependent photoluminescence measurements reveal the importance of the electron-tunneling-assisted nuclear spin pumping. We also find evidence for the resonant LO-phonon-mediated electron co-tunneling, the effect controlled by the applied bias and leading to the reduction of the nuclear spin pumping rate.  \\
\end{abstract}
\maketitle

The strong influence of the lattice nuclei on the electron spin properties in semiconductor nano-structures has been identified as a major challenge in the utilization of desirable electron spin properties for quantum logic applications \cite{Imamoglu,Koppens,Petta,Reilly,Atature}. This has sparked recent intense research efforts to control the nuclear spin in nano-structures, and particularly in III-V semiconductor quantum dots, where the hyperfine interaction between the magnetic moments of the electron and nuclear spins \cite{Overhauser} has been shown to limit the electron spin life-time \cite{Koppens,Erlingsson,Merkulov,Braun1,Oulton} and coherence \cite{Koppens,Petta,Reilly,Khaetskii}.

Recently, optical excitation \cite{Gammon,Bracker,Eble,Lai,Braun2,Maletinsky1,Maletinsky2,Tartakovskii,Urbaszek,Russell,Makhonin,Skiba} and transport \cite{Koppens,Petta,Koppens,Ono,Yusa} of spin-polarized electrons in semiconductor quantum dots (QD) have been shown to lead to dynamic nuclear polarization. The effect is a result of the hyperfine interaction leading to relaxation of electron spin via the "flip-flop" process, in which an electron transfers its spin to a single nucleus in the dot. In optically pumped dots Overhauser magnetic fields, $B_N$, up to a few Tesla have been generated under excitation with circularly polarized light \cite{Braun2,Maletinsky1,Tartakovskii,Skiba}. Strong non-linearities and noise associated with nuclear spin memory effects have been found in electron transport measurements on GaAs-based dots and quantum point contacts \cite{Koppens,Ono,Yusa1}. 

In this paper, we show that large Overhauser fields (up to 3 T) can be controlled in a single optically active dot grown in the intrinsic region of a semiconductor diode by applying small changes to the diode bias. Here the effect is observed in InGaAs/GaAs dots in the nuclear spin bi-stability regime in external field of $1.5<B_z<3$T \cite{Braun2,Maletinsky1,Tartakovskii,Urbaszek,Skiba}. We demonstrate bias-controlled switching between two distinctly different and stable collective nuclear spin states in a dot. A strong dependence of the optically pumped nuclear spin on the dot on the probability of the non-radiative escape of the electron from the dot to the back contact is demonstrated. In some bias regimes this process leads to increased nuclear polarization due to the electron-tunneling-mediated nuclear spin pumping: the photo-excited electron virtually occupies the inverted spin-state while remaining at the same energy, flops the spin of a single nucleus and tunnels out of the dot into a continuum of states in the contact. However, we also find that a resonant  phonon-assisted electron co-tunneling can occur, leading to depolarization of the electron on the dot and consequent lowering of the nuclear spin pumping rate.

We present the results for dots grown in the intrinsic region of an n-type Schottky diode. The device is grown by MBE on an undoped GaAs substrate. It consists of a 50 nm GaAs Si-doped back contact on top of which the following undoped layers are deposited: 25 nm GaAs tunnel barrier, InGaAs dots, 125 nm GaAs, 75 nm Al$_{0.3}$Ga$_{0.7}$As, 5 nm cap GaAs layer. The sample was then covered with an opaque metal mask, where sub-micron clear apertures were open for optical access to individual dots. In the micro-photoluminescence (PL) experiment presented here, circularly polarized optical excitation was employed to generate e-h pairs in 
the low energy tail of the wetting layer at 1.425eV. Unpolarized PL from the dots at $\approx1.32$eV was measured by a double spectrometer and a CCD. The identification of PL peaks employed here is carried out following the standard procedure for dots embedded in an n-type Schottky diode \cite{Warburton} and is based on the bias- and power-dependence of their intensities and spectral positions.

\begin{figure}[hb]
\centering
\includegraphics[width=8cm]{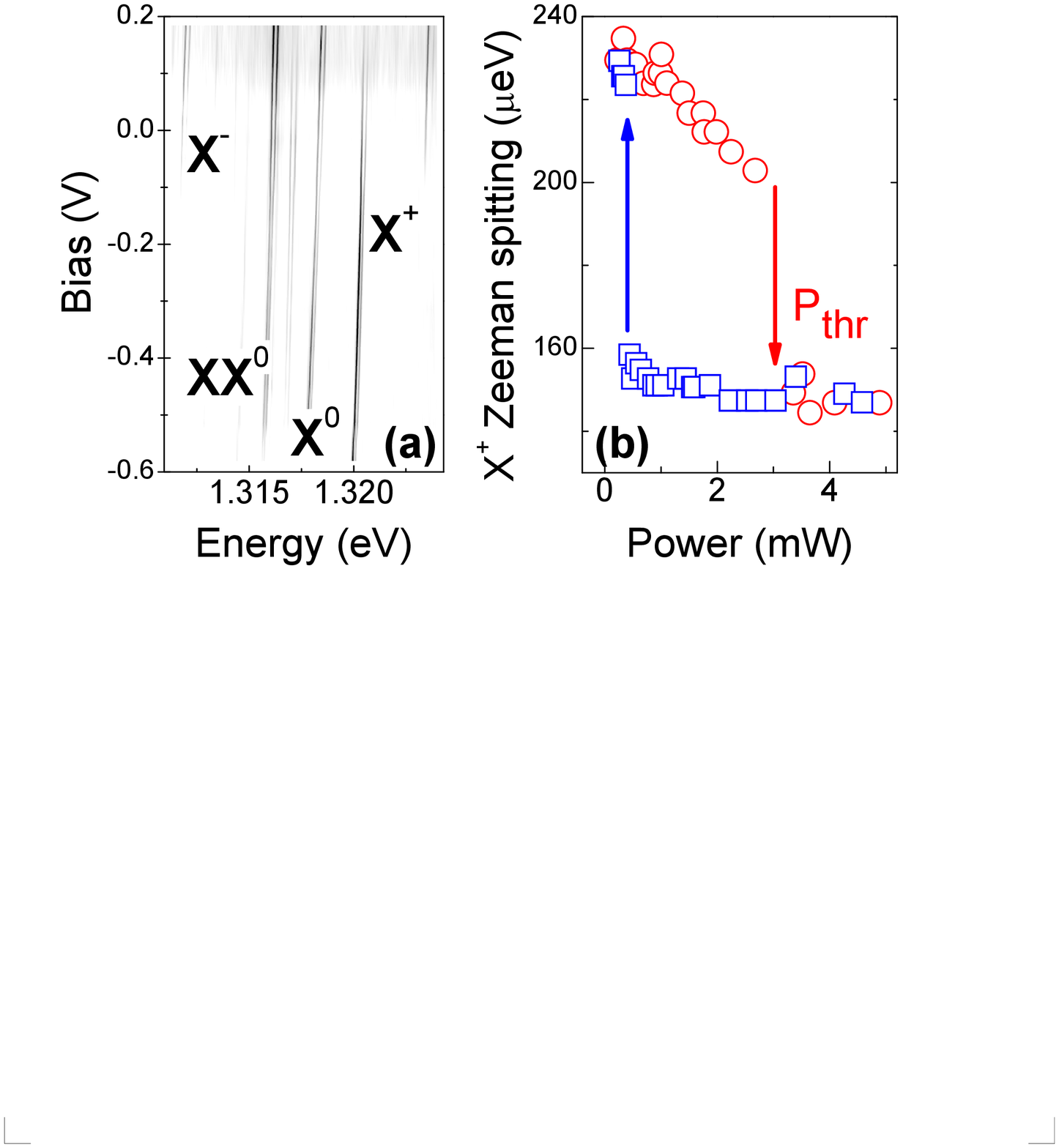}
\caption{(a) Bias-dependence of the QD PL recorded at $B_z=1.5$T under $\sigma ^{-}$ excitation. (b) Power dependence of the $X^+$ Zeeman splitting 
measured at $B_z=2$T and bias -0.45 V for $\sigma ^{-}$ polarized excitation.
$P_{thr}$ denotes the power where the nuclear spin ''switch'' is observed in the scan where the power is increased. The arrows at the switching thresholds indicate the directions of the power scans.}
\label{fig1} 
\end{figure}

Fig.1a shows bias-dependent PL spectra measured for high power (0.5 mW) laser excitation and external magnetic field $B_z=1.5$T. Observed PL arises from an e-h pair recombination when the following combination of charges is confined in the dot:  ehh (positively charged exciton, $X^+$), eh
(neutral exciton, $X^0$), eehh (bi-exciton, $XX$), eeh (negatively charged 
exciton, $X^-$). Note, that in the time averaged spectra measured in PL experiments, 
the relative intensities of the PL peaks represent probabilities to find the dot with a certain e-h population forming an exciton complex, which can recombine radiatively. 

As seen in Fig.1a at $B_z=1.5$T each of these exciton states exhibit Zeeman splitting. The electron spin is also sensitive to the effective magnetic (Overhauser) field $B_N$ produced by the polarized nuclei, so that the total magnetic field seen by the electron is ${\bf B}_{tot}={\bf B}_{z}+{\bf B}_N$. When circularly polarized excitation is employed to generate electrons with well-defined spin on the dot, nuclear spin polarization builds up as a result of the electron-to-nuclear spin transfer induced by the hyperfine interaction. ${\bf B}_N$ (anti)parallel to ${\bf B}_{z}$ is observed for ($\sigma^-$)$\sigma^+$ circularly polarized excitation. The electron Zeeman splitting can be written as
$E_e(\sigma^\pm) = g_e\mu_B(B_z \pm B_N)$, where $g_e$ is the electron g-factor and $\mu_B$ is the Bohr magneton. Note that the hyperfine interaction is negligible for holes due to their p-like wave function.

The efficiency of the electron-to-nuclei spin transfer, 
\begin{equation}\label{ws}
w_s\propto w_x|A_{hf}|^2/(\Delta E_e^2+\gamma^2/4),
\end{equation}
depends on the major energy cost of the electron-nuclear spin flip-flop event \cite{Erlingsson} - the electron Zeeman splitting $E_e$ \cite{Eble,Braun2,Maletinsky1,Tartakovskii,Urbaszek,book}. 
In this expression $w_x$ is the optical pumping rate, $A_{hf}$ is the hyperfine interaction constant and $\gamma$ is the electron state broadening \cite{Erlingsson,book}. The resonant form of the rate $w_s$ in Eq.\ref{ws} assumes a feedback between the nuclear spin polarization on the dot, influencing $\Delta E_e$, and $w_s$. In particular, an
abrupt build-up of nuclear spin on the dot under optical
excitation, the so-called  nuclear spin ''switch'' effect
\cite{Braun2,Maletinsky1,Tartakovskii,Urbaszek,Russell,Skiba}, occurs when $E_e\approx 0$ for  $B_N\approx -B_z$ leading to a sharp increase
of $w_s$.

An example of this phenomenon is shown in Fig.1b where the $X^+$ Zeeman splitting is plotted as a function of the pumping power for $\sigma^-$ polarized excitation, $B_z=2$T and bias -0.45V. As the power is increased from zero (circles), a gradual decrease in the $X^+$ Zeeman splitting is observed due to 
the increasing $B_N$ and hence decreasing $E_e$. At the power $P_{thr}=0.3$mW a threshold-like ''switching'' occurs due to a significant build up of the nuclear polarization and occurrence of high $B_N\approx B_z$. For powers ($P$) above $P_{thr}$ a weak power-dependence of the nuclear polarization is observed. If $P$ is reduced again, the switching to the low nuclear polarization branch occurs at $P<P_{thr}$, i.e. a hysteresis is observed constituting observation of the optically induced nuclear spin bistability. Note, that in Fig.1b the Overhauser field $B_N$ abruptly changes by more than 1T at the switching thresholds. In what follows we will consider in detail the dependence of the switching behavior in Fig.1b on the bias applied to the diode. We will show that the bias tuning produces dramatic changes in the nuclear polarization on the dot.

Fig.2a shows the $X^+$ Zeeman splitting $E_{X+}$ as a function of applied bias measured for a fixed power of 0.4mW of $\sigma^-$ polarized excitation at $B_z=2.1$T. The bias scanning directions are shown with arrows. The curve depicted with circles corresponds to the bias tuned from forward +0.2V to reverse -0.6V. A slight oscillation of the splitting around 240$\mu$eV is observed before a threshold-like transition to a notably lower $E_{X+}\approx170\mu$eV is detected at $V_{thr1}\approx-0.45$V. The transition corresponds to an abrupt build-up of nuclear polarization and occurrence of $B_N\approx2$T. When the bias is scanned in the opposite direction (shown with squares) the high magnitude of $B_N$ is observed up to $V_{thr2}\approx -0.1$V, where a threshold-like transition to a low magnitude of $B_N\approx0$ occurs. A pronounced hysteresis is clearly seen in Fig.2a. The observed threshold-like behavior of the Overhauser field indicates that large Overhauser fields in the Tesla range can be manipulated in a single dot by small changes of applied bias. The bias-controlled switching is observed up to the maximum $B_z=3$T \cite{Tartakovskii}, where the switching of $B_N\approx 3$T is achieved. The switching thresholds depend on the excitation power and magnetic field, since these two parameters define the nuclear spin pumping rate $w_s$ as seen from Eq.\ref{ws}: (i) a higher rate is achieved at a high excitation power due to the increased electron spin flux through the dot and (ii) $w_s$ is reduced at higher $B_z$ due to the increased $E_e$. 

We have carried out power-dependent measurements as in Fig.1b for a range of biases. It is observed that the nuclear spin switch can be achieved under $\sigma^-$ excitation in a wide range of voltages -0.6V$<V_{app}<$0V. The dependence of $P_{thr}$ on the bias (shown in Fig.2b) reveals a notable reduction of $P_{thr}$ at reverse biases below -0.4V.  This bias-dependence correlates well with the strong bias-dependence of $E_{X+}$ in Fig.2a: the switching of $E_{X+}$ to a lower magnitude when tuning the bias from +0.2V to -0.6V occurs at a bias where a notable reduction of $P_{thr}$ is measured, i.e. a lower pumping rate is required to efficiently polarize the nuclei. Note that a pronounced resonance-like feature with a notable increase of $P_{thr}$ is observed around  -0.3V in the bias-dependence of $P_{thr}$ (to be discussed below).

\begin{figure}[htb]
\centering
\includegraphics[width=8cm]{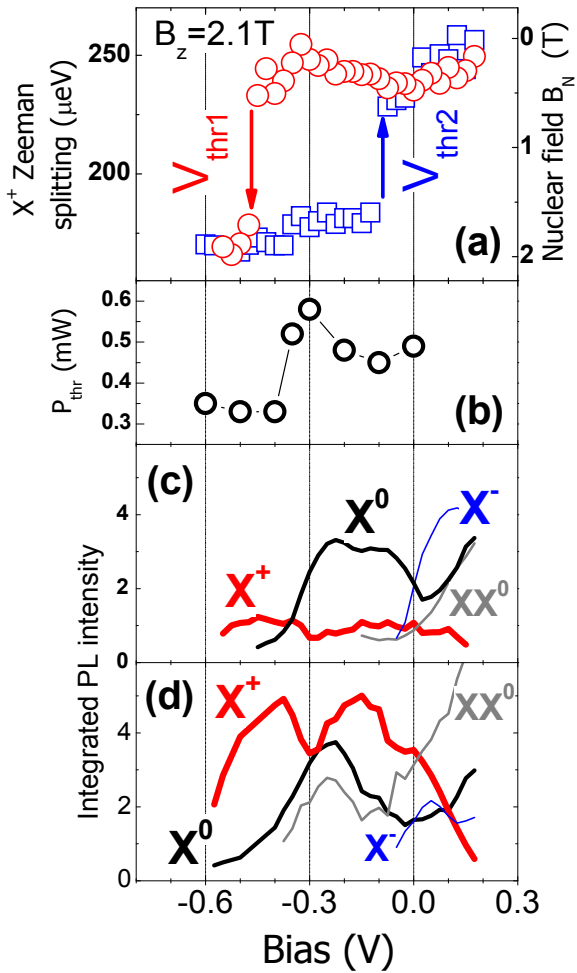}
\caption{(a) Voltage scans at $B_z=2.1$T and power 0.4 mW under $\sigma^-$ excitation. $V_{thr1}$ ($V_{thr2}$) denotes the bias where the nuclear spin switching occurs when reducing (increasing) the bias. Directions of the bias scans are shown with arrows. (b) The bias-dependence of the switching threshold power $P_{thr}$.
(c and (d) Bias-dependence of the integrated PL intensity measured in both circular polarizations at $B_z=2.1$T under $\sigma^-$ excitation with powers 0.2 (c) and 0.5 (d) mW for $X^0$ (electron and hole pair or eh state), $XX^0$ (eehh), $X^+$ (ehh) and $X^-$ (eeh) peaks.
}
\label{fig2} 
\end{figure}

In order to understand the observed bias-dependent behavior, we consider the electron dynamics on the dot that can be revealed by detailed PL measurements showed in Figs.3a,b. Here the integrated intensity of the unpolarized PL, measured under $\sigma^-$ excitation is plotted as a function of applied bias. Fig.2 reveals that both for powers above and below the switching threshold (Fig.2c and d, respectively) a notable quenching of PL signal occurs at biases below -0.4V. Separately conducted photo-current experiments on this sample, performed with the resonant excitation in the ground state of the dot, also showed that the onset of the PC signal, corresponding to the electron tunneling from the lowest state in the dot to the contact \cite{Oulton1,Beham}, occurs at $\approx -0.4$V \cite{Kolodka}. The range of biases below $-0.4$V coincides with the regime where the most efficient nuclear spin pumping on the dot occurs and a marked decrease in the threshold power for the spin switch is found.

The observed bias-sensitivity of the dynamic nuclear polarization can be understood from the mechanisms of the nuclear spin pumping in regimes with and without electron tunneling. In the PL regime the optically pumped carriers escape from the dot via radiative recombination.  When an e-h pair is excited in an unoccupied dot, the nuclear spin pumping occurs via initial formation of a dark exciton (with an electron preserving its spin and a depolarized hole) and a consequent virtual process involving electron-nuclear spin flip-flop and the dark exciton recombination \cite{Gammon,Tartakovskii}. In the case of the dot charging with a hole and $X^+$ formation, a similar spin-flip process is also possible, since the hole states with the two orientations of the spin are available for e-h recombination. In the case of the bi-exciton excitation the electron spin information is lost and the nuclear spin pumping is suppressed. Finally, in the regime of the dot negative charging the nuclear spin pumping occurs via the spin transfer from the resident electron remaining on the dot after the e-h recombination \cite{Lai,Eble,Maletinsky1,Maletinsky2,Skiba}. In Schottky structures this has been found to lead to $B_N$ anti-parallel to the Overhauser field pumped when a neutral and positively charged exciton is generated \cite{Eble,Maletinsky1}. 

We thus can expect a relatively high efficiency of the generation of the Overhauser field with the sign corresponding to the spin switch condition ($B_N$ anti-parallel $B_z$) in the range of biases where $X^+$ and $X^0$ are strong, but at the same time a weaker spin pumping effect at voltages where the contribution from $XX^0$ and $X^-$ is significant. In Fig.2 the $X^0/X^+$ regime spreads from 0V to -0.6V, beyond which PL is almost completely quenched. On the other hand, as seen from the bias-dependence measured at a high power (Fig.2d), $XX^0$ becomes strong at $V_{app}\approx -0.3$V and above 0V and $X^-$ appears at $V_{app}>0$V. This correlates well with the increasing threshold power for the spin switch at these voltages as depicted in Fig.2b, and can be explained by the detrimental effect of the formation of these e-h complexes on the nuclear spin pumping.

An additional nuclear spin pumping mechanism becomes possible in the regime where the non-radiative escape of the electron from the lowest energy dot states to the contact becomes possible due to the tunneling at high applied voltages $V_{app}\leq-0.4$V. This mechanism has been recently proposed by us in Ref.\cite{Makhonin1}: the photo-excited electron virtually occupies the inverted spin-state while remaining at the same energy, flops the spin of a single nucleus and tunnels out of the dot into a continuum of states in the contact. In addition, as seen from Eq.\ref{ws}, the degree of nuclear spin polarization is dependent on the rate of re-excitation of the dot with a spin polarized electron \cite{Tartakovskii}. In the low voltage regime this is limited by the rate of the e-h radiative recombination, whereas in the regime of high bias, an additional fast electron escape route appears due to the tunneling. In contrast to the case of the resonant excitation considered in Ref.\cite{Makhonin1}, the re-excitation of the spin-polarized electrons on the dot by the non-resonant optical pumping is not conditional on the fast hole escape. Thus the electron re-pumping rate and, as a consequence, the nuclear spin pumping can be considerably enhanced in the regime of high voltage compared to the range of $V_{app}$, where the radiative recombination dominates and clear evidence of the complete blocking of the dot ground state (observation of $XX^0$) is found.

The same factors influencing the nuclear spin pumping on the dot will determine the magnitude of $V_{thr2}$, where the switching back to a low nuclear polarization is observed. In this case, however, a high nuclear spin pumping rate can be maintained far into the regime of radiative recombination due to the initial cancellation of $E_{e}$ by the Overhauser field at high bias (see Eq.\ref{ws}).  

We now discuss the resonant feature observed at $\approx -0.3$V in the bias-dependence of the switching threshold power $P_{thr}$ shown in Fig.2b. Firstly, we note that the efficient electron charging arising from the tunneling of the electrons $from$ the contact into the dot, occurs at $V_{app}\approx 0$V. This bias corresponds to the lining-up of the electron ground state in the dot and the edge of the Fermi sea in the contact. Correspondingly, when a reverse bias $V_{app}$ is applied, the electron ground state is about $\Delta E_{GS}=e|V_{app}|d_{bar}/d_{tot}$ above the edge of the Fermi sea, where $d_{bar}$ and $d_{tot}$ are the thickness of the tunneling barrier below the dot and the total thickness of the intrinsic region, respectively. For $V_{app}=-0.3$V, $d_{bar}=25$ nm and $d_{tot}=230$ nm, we obtain $\Delta E_{GS}=33m$eV, very close to the energy of the GaAs optical phonon. 

We suggest that near $V_{app}=-0.3$V the following process, best described as a phonon-assisted resonant co-tunneling \cite{Smith}, can take place: (i) the electron from the dot tunnels to the contact with an emission of an optical phonon; (ii) this phonon is absorbed by another electron from the contact, which then tunnels into the ground state of the dot. As in the case of co-tunneling previously reported for charge-tunable devices \cite{Lai,Maletinsky2,Smith}, this process can lead to the electron spin relaxation. This will result in lowering of the pumping rate of the nuclear spin on the dot, as indeed observed from the increase of the spin switch threshold in Fig.2b. In addition, this spin depolarization mechanism will increase the probability of the electron spin singlet formation in the optically pumped dot, otherwise inefficient due to a rather low probability of the electron spin flip during relaxation from the wetting layer to the ground state of the dot. This gives rise to the increased probability to form bi-excitons (and as a result neutral excitons) and decreased probability to form $X^+$, which again is observed at $V_{app}\approx -0.3$V in Fig.2d. All observed effects, the increase of $P_{thr}$, increase in $XX^0$ and $X^0$ PL and decrease in $X^+$ PL occur in a narrow range of biases around $V_{app}\approx -0.3$V, which supports our suggestion about the resonant nature of the observed phenomenon, namely the phonon-assisted electron co-tunneling.  

In conclusion, we found an efficient way of controlling nuclear spin in an individual quantum dot by applying small changes in the electric field across the device. This can serve as an efficient technique in controlling the magnetic environment of the electron spin on the dot by employing an externally controlled high-rate switching of the Overhauser field in the range of several Tesla. This will be achieved by applying fast bias bursts combined with ms-long optical pulses \cite{Maletinsky2,Makhonin} sufficient to pump high nuclear spin polarization in an individual dot.  

This work has been supported by the EPSRC grants GR/S76076, EP/G601642/1, the EPSRC IRC for Quantum Information Processing, and the Royal Society. AIT acknowledges support from the EPSRC grants EP/C54563X/1, EP/C545648/1.

\pagebreak
\end{document}